\documentclass[amsmath,amssymb,prl]{revtex4}
\usepackage{dcolumn}
\usepackage{graphicx}
\begin{document}
\title{Reversing non-local transport through a superconductor 
by electromagnetic excitations}
\author{A. Levy Yeyati, F.S. Bergeret and A. Mart\'{\i}n-Rodero}
\affiliation{Departamento de F\'{i}sica Te\'{o}rica de la Materia
Condensada C-V, Universidad Aut\'{o}noma de Madrid, E-28049
Madrid, Spain}
\author{T.M. Klapwijk}
\affiliation{Kavli Institute of Nanoscience, Delft University of
Technology, Lorentzweg 1, 2628 CJ Delft, The Netherlands}
\begin{abstract}
Superconductors connected to normal metallic electrodes at the nanoscale
provide a potential source of non-locally entangled electron pairs. Such 
states would arise from Cooper pairs splitting into two electrons with 
opposite spins tunnelling into different leads. In an actual system the 
detection of these processes is hindered by the elastic transmission 
of individual electrons between the leads, yielding an opposite contribution 
to the non-local conductance.  Here we show that electromagnetic excitations 
on the superconductor can play an important role in altering the balance 
between these two processes, leading to a dominance of one upon the other 
depending on the spatial symmetry of these excitations. These findings 
allow to understand some intriguing recent experimental results and 
open the possibility to control non-local transport through a superconductor 
by an appropriate design of the experimental geometry. 
\end{abstract}
\maketitle

In 1964 A.F. Andreev proposed a mechanism to explain the conversion of 
quasiparticle currents into supercurrents at the interface between a normal 
metal (N) and a superconductor (S) \cite{andreev}. In the so-called Andreev 
reflection mechanism an incident electron from the N region is reflected as 
a hole and a Cooper pair is created on the superconductor. This process can 
be viewed as the transfer of two electrons with opposite spins from N to S
where they combine to form a Cooper pair. In multiterminal N/S structures
a non-local version of this process where the electrons are injected
from two spatially separated electrodes, can take place 
\cite{flatte,hartog,deustcher}. 
The possibility to detect these {\it crossed Andreev reflection} (CAR) 
processes have attracted a lot of interest since it provides a natural 
mechanism to produce non-locally entangled electron pairs in a condensed 
matter device \cite{recher,chtchelkatchev,bena,samuelsson,prada}. 
Unfortunately, a competing mechanism 
known as elastic cotunnelling (EC) in which electrons are transmitted
elastically between the two electrodes, is always present in actual
devices yielding an opposite contribution to the non-local conductance.
The theory, in fact, predicts that for BCS superconductors weakly coupled
to non-magnetic leads the contributions of EC and CAR processes to the
non-local conductance tend to cancel each other \cite{falci,feinberg}. 
Spin polarisation due to magnetism in the leads could break this 
cancellation \cite{falci,melin}, an idea that has been explored in 
recent experiments \cite{beckmann}.
Surprisingly, a different set of experiments by Russo et al. \cite{russo} 
has shown that even for the case of non-magnetic leads coupled to the 
superconductor by tunnel junctions the subgap non-local conductance can be 
appreciably large, exhibiting an intriguing behaviour in which either
process can dominate depending on the applied voltage range.
 
\begin{figure}
\includegraphics[scale=0.5]{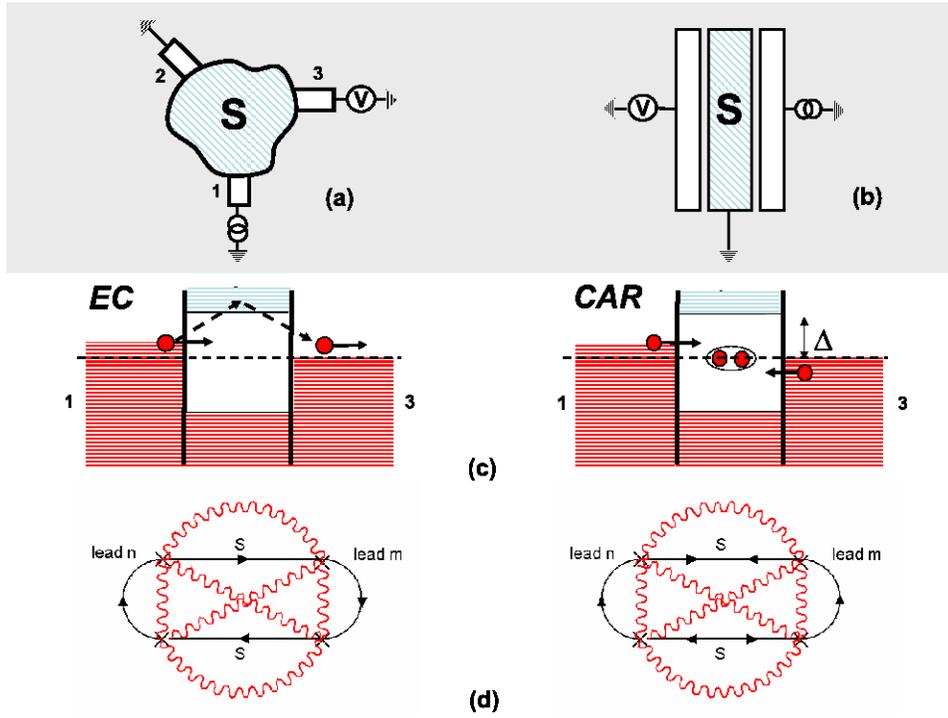}
\caption{Top panel: a) Schematic representation of a generic 
multiterminal geometry where a superconducting region is coupled
to several metallic leads and b) Double planar N/S/N junction geometry studied
in Ref. \cite{russo}. 
Middle panel (c): Pictorial description of EC and CAR processes in
energy space. Lower panel (d): Feynman diagrams corresponding to the
calculation of the non-local conductance to fourth order taking into
account interactions mediated by the electromagnetic environment. Full lines
with an arrow represent the normal and anomalous propagators and
wavy lines indicate phase correlators.}\label{geometry}
\end{figure}

In this work we analyse the influence of electron interactions in the
transport properties of this type of multiterminal N/S structures. 
A key feature of this analysis is played by low energy collective 
excitations which appear in superconductors of reduced dimensionality
\cite{mooij}. When an electron tunnels into the superconductor it
can excite such low energy modes which alters the balance between EC and
CAR processes. Moreover, depending on the spatial symmetry of these 
excitations either CAR or EC processes can be suppressed, leading to a change 
in the sign of the non-local conductance as a function of the applied 
voltage. These findings provide an explanation to the experimental results 
of  Ref. \cite{russo} and suggest new strategies for the detection of CAR 
induced transport in nanostructures.

A generic set-up for measuring the non-local resistance is shown in 
Fig.\ref{geometry}a. 
It represents a superconducting region attached to three normal 
electrodes. Two of the leads (labelled 1 and 2 in Fig.\ref{geometry}a)
are used to inject a current while the 
voltage drop is measured on the third one. As illustrated in 
Fig.\ref{geometry}c, EC processes produce and injection of electrons
into the third lead while CAR processes withdraw them from it, 
which yields opposite contributions to the non-local resistance 
\cite{comment}.
On the other hand both processes decay on the superconducting coherence 
length $\xi$, which can range between 10 and 100 $nm$ for typical 
superconductors used in experiments \cite{russo,beckmann}. 

The importance of interactions in breaking the balance
between EC and CAR processes can be understood by considering the case 
where the S region is sufficiently small and can be characterised by 
a finite charging energy $E_c$.
As it is sketched in Fig.\ref{geometry}c, EC processes take place through  
a virtual state which will be shifted upwards by the Coulomb energy. 
The process, however, would not be blocked for any value of the
applied voltage as the initial and final states have the
same energy. In contrast, CAR processes demand that two electrons 
tunnel into the S region forming a Cooper pair, a process which
requires an extra energy of $4E_c$. Thus, 
the non-local conductance has a finite (negative) value for a
voltage $V$ smaller than $4E_c/e$ where EC processes dominate, while 
it vanishes for $eV>4E_c$ when both processes cancel each other. 

For a quantitative analysis of the influence of interactions we describe
the system by a Hamiltonian $\hat{H} = \hat{H}_s + \hat{H}_{leads} + 
\hat{H}_T+ \hat{H}_{env}$.
The first three terms correspond to the electronic degrees of freedom. 
$\hat{H}_S$ is the usual BCS Hamiltonian for the S region and 
$\hat{H}_{leads}$ describes the normal leads which we label with an index
$n$.  The tunnelling of electrons between the leads and the superconductor 
is described by $\hat{H}_T = \sum_{n} \hat{H}_{T,n}$, with 
\begin{equation}
\hat{H}_{T,n} = \sum_{\sigma} \int_{S_n} d^2r \left[ v_{n} 
\hat{\psi}^{\dagger}_{ln,\sigma}(\vec{r}) \hat{\psi}_{sn,\sigma}(\vec{r}) 
e^{i \hat{\phi}_n(\vec{r})} \; \; + \mbox{h.c.} \right], 
\end{equation}
where the integral is taken over the junction area $S_n$, 
$\hat{\psi}^{\dagger}_{ln,\sigma}(\vec{r})$ and 
$\hat{\psi}^{\dagger}_{sn,\sigma}(\vec{r})$ 
are electron creation operators on the two sides of junction
and $\hat{\phi}_n(\vec{r})$ is the corresponding phase drop
which is conjugate to the charge density on the junction 
$\hat{Q}_n(\vec{r})$, i.e $[\hat{\phi}_n(\vec{r}),Q_n(\vec{r}')]
=ie\delta(\vec{r}-\vec{r}')$.  
The dynamics of these phase operators is determined by the Hamiltonian
$\hat{H}_{env}$ describing the electromagnetic environment characterising
the actual experimental set-up.
It is important to note that, due to the typical distances between
the leads in the experiments, which are not much larger than $\xi$, 
correlations between voltage fluctuations on different junctions
cannot be neglected, i.e. correlation functions of the type
$<\hat{\phi}_n\hat{\phi}_m>$, with $n\neq m$ are non-zero. 
In addition, the reduced dimensions of the S region can give
rise to the presence of collective modes within the superconducting
gap, which can dominate the behaviour of the phase
correlations.

\begin{figure}
\includegraphics[scale=0.5]{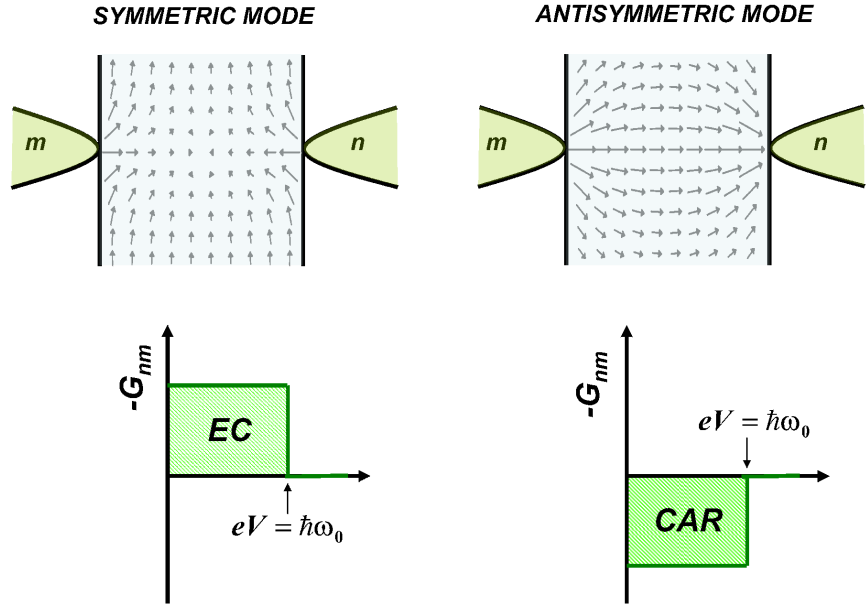}
\caption{Pictorial representation of the effect of interactions mediated
by electromagnetic modes of different symmetry on the non-local conductance 
between two point contacts. The arrows represent the phase gradient 
within the superconductor. While symmetric modes tend 
to suppress CAR processes the antisymmetric ones suppress the EC
contribution.}\label{modes}
\end{figure}

To obtain the transport properties of this model we use a combination of
Keldysh and Nambu formalisms which is well adapted to analyse non-equilibrium 
situations in the presence of superconductivity \cite{methods}.
The contributions from EC and CAR processes to the non-local conductance
$G_{nm}$,
i.e. the variation of the current through lead $n$ due to a voltage
applied on lead $m$, in 
the tunnel limit is represented by the type of diagrams   
shown in Fig.\ref{geometry}d. The solid lines with an arrow represent the
electron propagators, while the wavy lines describe the coupling with the 
environment, i.e. they denote the phase correlators of the type
$<e^{i\hat{\phi}_n} e^{-i\hat{\phi}_m}>$. 
Let us first consider the simplest case where the environment can be
characterised by a single electromagnetic mode of frequency $\omega_0$.
One can further assume that the leads are coupled to the S region 
through point contacts as illustrated in Fig.\ref{modes}. 
Two opposite situations can be distinguished depending on 
the spatial symmetry of the electromagnetic mode under consideration: it can 
lead to either symmetric or antisymmetric voltage fluctuations on
the two junctions. 
In the symmetric case, assuming that $\hbar\omega_0$ is much smaller than
the superconducting gap $\Delta$, 
and for S in the clean limit, at zero temperature
we obtain (see Supplementary Information)
\begin{eqnarray}
G_{nm} &=& -\frac{G_n G_m}{G_0} 
\frac{e^{-2R/\xi}}{(k_FR)^2} e^{-2z_0}
\sum_{n_1...n_6}\left[ \prod_{i=1}^6 \frac{(z_0)^{n_i}}{n_i!} \right]
\theta(eV-\sum_{i=1}^4 n_i\hbar\omega_0)
\left[(-1)^{n_3+n_4}\cos^2(k_FR) - (-1)^{n_5+n_6}
\sin^2(k_FR) \right] \label{gnls}. \nonumber \\
\end{eqnarray}

Here $G_{n(m)}$ is the normal conductance of the junction $n$ ($m$), $R$  
the distance separating the leads, $k_F$ is the Fermi wavevector, 
$G_0 = 2e^2/h$ is the conductance quantum and $V$ 
is the voltage applied on lead $m$. 
The term proportional to $\cos^2(k_FR)$ corresponds to the EC 
contribution while the $\sin^2(k_FR)$ term arises from CAR processes. 
The parameter $z_0=E_c/\hbar\omega_0$, where $E_c=e^2/2C$ is the charging
energy on the tunnel junctions, measures the coupling to the
electromagnetic mode. The indexes $n_i$, with $i$ ranging
from 1 to 6 denote the number of quanta associated to the phase correlators 
on each diagram in Fig.\ref{geometry}d, 5 and 6 corresponding to the more 
external wavy lines. The behaviour predicted by Eq. (\ref{gnls}) 
becomes more clear in the $z_0 \ll 1$ limit where it
can be simplified to obtain 

\begin{equation}
G_{nm} \simeq -\frac{G_n G_m}{G_0} 
\frac{e^{-2R/\xi}}{(k_FR)^2}
\left[\cos^2(k_FR) - 
(1 - z_0 \theta(\hbar\omega_0-eV)) \sin^2(k_FR)\right]
\; .
\end{equation}

It is worth noticing that this expression reproduces the non-interacting
result \cite{falci} for $z_0 = 0$, where a complete cancellation between
CAR and EC contributions takes place upon averaging over the Fermi 
wavelength scale. For finite but small $z_0$ the balance between EC and
CAR is broken:
for $eV$ smaller than $\hbar\omega_0$ the 
CAR processes become suppressed and non-local transport is dominated by the
EC contribution, while for $eV>\hbar\omega_0$ both 
contributions tend to cancel, as in the non-interacting case.  
The suppression of the CAR contribution is due to the impossibility of 
such processes to occur without producing a real excitation of the 
environment, as in the constant charging energy example.

The situation is the opposite in the case of an antisymmetric mode. 
The analog of expression to Eq. (\ref{gnls}) for this case is 
obtained by an exchange of the indexes $n_3,n_4$ with $n_5,n_6$
in the last factor within brackets. 
As a consequence EC (instead of CAR) processes will be suppressed 
at low voltages. 
The different effect of symmetric and antisymmetric modes is schematically
illustrated in Fig.\ref{modes}. 

We now consider the case of a planar geometry similar to the one in 
the experiments of Ref. \cite{russo} (Fig.\ref{geometry}b), consisting of 
a superconducting 
layer of thickness $d\gtrsim \xi$ coupled to two normal leads by
tunnel junctions.  The cross section of the junctions in Ref. \cite{russo}
was relatively large ($\sim 30 \, \mu m^2$), which allows to describe
them as infinite planes \cite{pierre}.
The electromagnetic environment in this situation is characterised by the
presence of propagating modes along the S/N junctions, which 
can be derived from the following model Hamiltonian
\begin{equation} 
\hat{H}_{env} = 
\left(\frac{\hbar}{2e}\right)^2
\int d^3r 
\frac{(\nabla \hat{\phi})^2}{2Ld} + 
\int d^2r \frac{1}{2C_{\square}} \left(\hat{Q}_L^2 + \hat{Q}_R^2\right) \, ,
\label{henv}\end{equation}
where the term containing the phase gradient describes the kinetic energy
associated to the supercurrents in the S film, $L$ being its total inductance,
while the second term is the Coulomb energy of the charge accumulated on
the junctions (assumed to be symmetric with capacitance $C_{\square}$ per unit area
and with cross section ${\cal S}$).
In writing this Hamiltonian we are assuming that long range Coulomb 
interactions are screened by the normal electrodes acting 
as ground planes \cite{fisher}. 
The low energy modes which result from this model correspond to 
symmetric and antisymmetric voltage fluctuations on the junctions,
with dispersion relations $\omega_1(\vec{q}) = c_s \sqrt{(q \tanh{qd})/d}$
and $\omega_2(\vec{q}) = c_s \sqrt{q/(d \tanh{qd})}$, where $\vec{q}$
is the wavevector in the direction parallel to the film and 
$c_s = 1/\sqrt{LC_{\square}}$. Notice that for small $q$ the symmetric mode exhibits 
a linear dispersion with phase velocity $c_s$ while
the antisymmetric one tends to a finite frequency $\omega_0 = c_s/d$
in the limit $q \rightarrow 0$. This description of the low energy modes
captures the essential features of a detailed calculation based on 
Maxwell equations for the double planar junction geometry
(see Supplementary information).

One can roughly estimate the order of magnitude of the parameters in 
$\hat{H}_{env}$ for the experimental situation. Thus, $C_{\square}$
can be obtained from the typical charging energy for an oxide barrier
tunnel junction
$E_c{\cal S} \sim 1 \mu eV \times \mu m^2$ and  
$L$ can be estimated as $\mu_0 \lambda^2/d $,
$\lambda$ being the field penetration depth \cite{orlando}. 
The actual value of $\lambda$ for a Nb film is strongly dependent on
its thickness, degree of disorder and it is also influenced by 
the properties of the non-superconducting substrate on which it is
deposited \cite{gubin}. Reported values range between 100 $nm$ and 
1 $\mu m$ for $d \sim 10-100 \; nm$ 
\cite{gubin,parage}. Within this range of parameters the lowest 
energy of the antisymmetric mode $\hbar \omega_0$ 
can be of the order of a few $meV$, i.e. comparable to
the superconducting gap in Nb, even for the smaller film thickness 
analysed in Ref. \cite{russo}.

To obtain the non-local conductance $G_{LR}$ measured at the left interface 
when a voltage $V$ is applied on the right junction we extend the theory
developed for the single mode case, linearising with respect to  
the coupling parameters $z_{1,2}(\vec{q}) = E_c/\hbar\omega_{1,2}(\vec{q})$,
which is justified for the range of parameters estimated above. 
We thus obtain 

\begin{eqnarray}
G_{LR} &=& 4\pi\frac{G_L G_R}{{\cal S}k_F^2 G_0} \; \mbox{Ei}(-2d/\xi)
\sum_{\vec{q},\alpha=1,2} 
(-1)^{\alpha}z_{\alpha}(\vec{q}) 
\left[\left(N(\omega_{\alpha}(\vec{q})+1\right) 
\left(F(\omega_\alpha(\vec{q}))+1 \right)  
+ N(\omega_{\alpha}(q)) 
\left(F(-\omega_\alpha(q))+1 \right)
\right], \label{gnl}
\end{eqnarray}

where $G_{L,R}$ is the normal conductance of each junction, 
$F(\epsilon)=\int d\omega \frac{\partial f(\omega)}{\partial\omega}[f(\omega+\epsilon-V)+f(\omega+\epsilon+V)]$ 
is a thermal smearing kernel arsing from
the Fermi distribution $f(\omega)$ while
$N(\omega_\alpha)$ is the Bose distribution function. The planar geometry 
leads to the factor
Ei(-$2d/\xi$), where Ei denotes the exponential integral function.
This leads to an exponential decay of the
non-local conductance when increasing $d$. 

\begin{figure}[ht]
\includegraphics[scale=0.4]{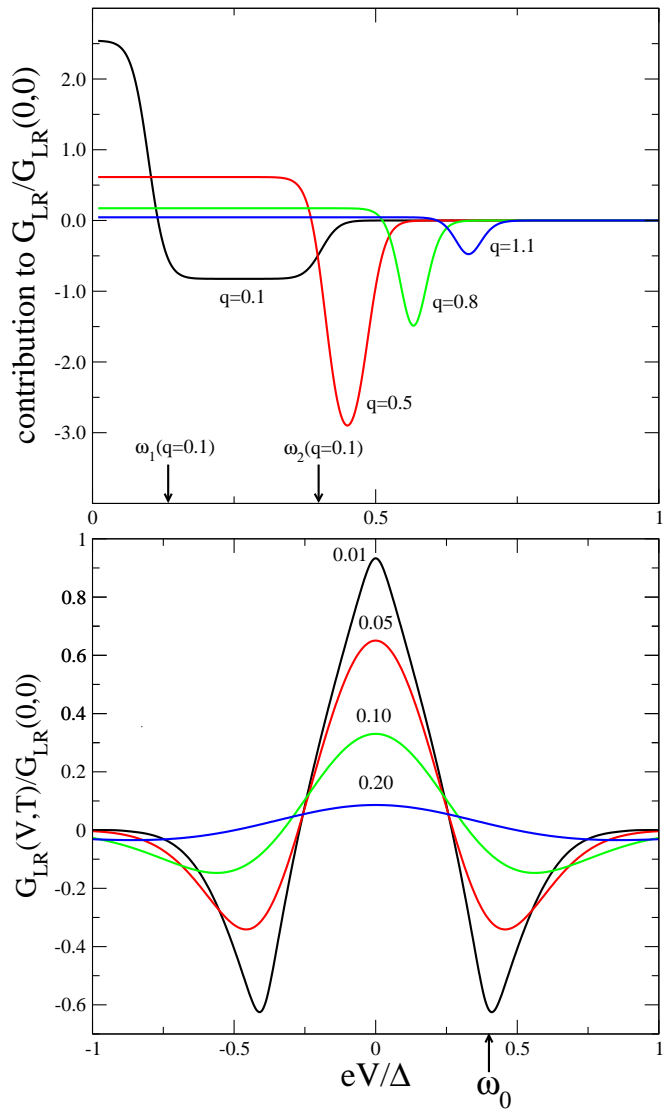}
\caption{Upper panel: Contribution to the non-local conductance from modes
with a given wavevector in the double planar junction geometry 
for $k_BT = 0.01 \Delta$ and 
$\hbar \omega_0 = 0.4 \Delta$. The values
of $q$ are given in units of $\Delta/\hbar c_s$. The arrows indicate
the energy of symmetric and antisymmetric modes for $q=0.1$. 
Lower panel: Temperature and voltage dependence of the total non-local
conductance. The temperature values are given in units of $\Delta/k_B$.
The arrow indicates the energy $\hbar\omega_0$ for the lowest
antisymmetric mode.}
\label{delft}
\end{figure}

In order to understand the behaviour of $G_{LR}$ as a function of voltage it 
is convenient to first analyse the contribution arising from a given 
wavevector $q$. 
This is illustrated in the upper panel of Fig.\ref{delft} 
for $T=10^{-2}\Delta$ and $\hbar\omega_0=0.4 \Delta$. 
The behaviour for the different wavectors is qualitatively similar:
for $eV<\hbar\omega_1,\hbar\omega_2$ the EC processes dominate, while CAR 
become more 
important in the voltage window $\hbar\omega_1<eV<\hbar\omega_2$ and 
finally both contributions cancel for $eV > \hbar\omega_2$.

The sum of all contributions yields a non-local conductance which is
dominated by EC processes at $V \rightarrow 0$ and decreases
almost linearly with $V$ until $eV \simeq \hbar\omega_0$. At this point the
onset of antisymmetric modes produces a strong suppression of EC processes
and there is a change of sign in $G_{LR}$ (CAR dominates). We can 
therefore associate $\hbar \omega_0$ with the crossing energy from
EC to CAR dominated regimes.
The lower panel of  Fig.\ref{delft} shows the voltage dependence 
of $G_{LR}$ for different temperatures. As can be observed, 
the imbalance between EC and CAR processes driven by the electromagnetic 
modes is less pronounced for increasing temperature.
The characteristic temperature for the suppression of the non-local
conductance is set by $\hbar \omega_0/k_B$.
This behaviour is in qualitative agreement with the results of 
Ref.\cite{russo}.
Moreover, the magnitude of the non-local conductance
predicted by our model is in reasonable agreement with the experimental
values. For instance, the ratio $G_{LR}/G^s_{L,R}$ between the non-local
and the direct conductances in the superconducting state at zero voltage
and zero temperature is 
$0.75 \, \mbox{Ei}(-2d/\xi) (E_c {\cal S} \omega_0/\hbar c_s^2)$ 
which yields values $\sim 10^{-3}$ 
close to the experimental ones for the parameters estimated above.  
A more quantitative description of
the experimental results will be presented elsewhere.

In conclusion we have shown that electron interactions 
mediated by electromagnetic excitations lead to an imbalance between 
EC and CAR processes.
Electromagnetic modes can either suppress CAR or EC processes depending
on their spatial symmetry. Taking into account that these low
energy excitations are strongly dependent on the geometrical 
characteristics of the multiterminal device,
these findings open the possibility to
control non-local transport processes through a superconductor by
an appropriate design of the experimental set-up.
For instance, one possibility would be to introduce an additional
tunnel junction inside the superconducting film in the experimental
geometry of Ref. \cite{russo}. This N/S/S/N nanostructure would allow
to control the dispersion relation of the electromagnetic modes
by varying the Josephson coupling between the S layers by means of a
magnetic field.
Let us finally point out that the high sensitivity of non-local transport 
to the electromagnetic modes could be used as a tool to analyse these 
excitations in hybrid nanostructures. 

\acknowledgments
The authors would like to thank fruitful discussions and correspondence with 
D. Beckmann, A. Morpurgo, S. Russo, C. Urbina, D. Esteve, W. Herrera, 
R.C. Monreal and J.C. Cuevas.


\begin{thebibliography}{25}
\bibitem{comment} Conventionally the CAR contribution to the non-local 
conductance is taken as positive, which corresponds to a negative contribution
to the non-local resistance. See Ref. \cite{falci}.
\end{thebibliography}
\end{document}